\documentclass[9pt,journal]{IEEEtran}

\usepackage{kpfonts}
\usepackage{seqsplit}
\usepackage{placeins}
\usepackage{booktabs}
\usepackage[labelfont=bf]{caption}
\usepackage{nameref}
\usepackage{rotating}
\usepackage{float}
\usepackage{graphicx}
\usepackage{multirow}
\usepackage{newfloat}
\usepackage{hyperref}
\usepackage{vcell}
\usepackage{enumerate}
\hypersetup{colorlinks,citecolor=blue,linkcolor=blue,urlcolor=blue}
\usepackage{array}
\newcolumntype{R}[1]{>{\raggedright\arraybackslash}p{#1}}
\newcolumntype{C}[1]{>{\centering\let\newline\\\arraybackslash\hspace{0pt}}m{#1}}
\usepackage{setspace}
\usepackage{amsmath}
\usepackage{algorithm}
\usepackage{algorithmic}
\usepackage{subcaption}
\usepackage{multicol}
\usepackage{xcolor}
\usepackage{soul}
\usepackage{xspace}
\definecolor{lightgreen}{rgb}{0.7, 0.9, 0.7} 
\sethlcolor{lightgreen}

\newcommand{\gDel}{gDel\_minRN\xspace}
\newcommand{\proposed}{DBgDel\xspace}

\begin{document}

\title{DBgDel: Database-Enhanced Gene Deletion Framework for Growth-Coupled Production in Genome-Scale Metabolic Models}

\author{

\IEEEauthorblockN{Ziwei Yang$^{1}$, Takeyuki Tamura$^{1}$}\\
\IEEEauthorblockA{\textit{$^{1}$ Bioinformatics Center, Institute for Chemical Research, Kyoto University}, Kyoto, Japan} 

\thanks{}}

\markboth{}%
{Shell \MakeLowercase{\textit{et al.}}: A Sample Article Using IEEEtran.cls for IEEE Journals}

\maketitle
\begin{abstract}
When simulating metabolite productions with genome-scale constraint-based metabolic models, gene deletion strategies are necessary to achieve growth-coupled production, which means cell growth and target metabolite production occur simultaneously. 
Since obtaining gene deletion strategies for large genome-scale models suffers from significant computational time, it is necessary to develop methods to mitigate this computational burden. 
In this study, we introduce a novel framework for computing gene deletion strategies. 
The proposed framework first mines related databases to extract prior information about gene deletions for growth-coupled production.
It then integrates the extracted information with downstream algorithms to narrow down the algorithmic search space, resulting in highly efficient calculations on genome-scale models. 
Computational experiment results demonstrated that our framework can compute stoichiometrically feasible gene deletion strategies for numerous target metabolites, showcasing a noteworthy improvement in computational efficiency. 
Specifically, our framework achieves an average 6.1-fold acceleration in computational speed compared to existing methods while maintaining a respectable success rate.
The source code of \proposed with examples are available on {{https://github.com/MetNetComp/DBgDel}}.

\end{abstract}


\section{INTRODUCTION}
\label{sec:introduction}
Computational approaches are essential in many metabolic engineering applications \cite{otero2021automated, garcia2021metabolic, foster2021building, toya2013flux,pharkya2006optimization}. 
A representative example is computational strain design, which relies on mathematical models to simulate a microorganism’s metabolic processes and the production of target metabolites. 
In genome-scale metabolic engineering simulations, the \textbf{constraint-based model} is one of the most popular models. 
This model typically comprises two components: (1) a metabolic network and (2) \textbf{gene-protein-reaction (GPR) rules}. 
A metabolic network serves as the backbone of the whole model. 
It contains chemical reactions that define the transformation relationships between metabolites.
Enzymatic proteins encoded by genes catalyze the chemical reactions within cells, associating these reactions with specific genes. 
Accordingly, GPR rules employ Boolean functions to represent these relationships between genes, proteins, and reactions. 
In this manner, a metabolic network and GPR rules can model the metabolic mechanisms of a specific microorganism.
Furthermore, a metabolic network can be modulated by influencing its reactions. 
Specifically, we can "turn off" certain reactions by deleting genes that encode the required enzymes.
This operation of deleting genes to reshape metabolic networks is the core idea of metabolic engineering.

During the simulation of gene deletion in constraint-based models, the primary objective is to achieve \textbf{growth-coupled production} for target metabolites. 
Growth-coupled production means cell growth is bonded to the synthesis of target metabolites in the metabolic process of microorganisms.
Specifically, in this study, we adopt the paradigm of weakly growth-coupled production, which means achieving a non-zero synthesis rate of the target metabolite at the maximum non-zero growth rate \cite{schneider2021systematizing, alter2019determination}.
Coupling cell growth with the production of target metabolites is necessary.
The reason is in industrial practices, microorganism genotypes with higher cell growth are more likely to persist in the culture through repeated passaging.
However, in the natural metabolic state of most microorganisms, only a limited number of metabolites meet the criteria for growth-coupled production. 
Therefore, achieving growth-coupled production for most target metabolites requires calculating gene deletion strategies to reshape the metabolic network \cite{vieira2019comparison}. 
However, gene deletion strategies calculation tasks are far from trivial.
Calculating gene deletion strategies demands huge computational resources, especially when simulating genome-scale models with complex metabolic networks and GPR rules involving many genes.

Several methods have been proposed to address this issue. 
Among the existing approaches \cite{pharkya2004optstrain,ranganathan2010optforce,burgard2003optknock,lun2009large,rockwell2013redirector,yang2011emilio,egen2012truncated,lewis2010omic,gu2016idealknock,ohno2014fastpros,tamura2018grid} for calculating gene deletion strategies in growth-coupled production, the most efficient ones include the minimal cut set (MCS)-based method that originated from the elementary flux vector (EFV)-based method \cite{von2017growth}. 
The basic idea of the EFV method is to identify a minimal set of reactions in the flow where cell growth compels the production of the target metabolite. 
Specifically, it determines a non-decomposable flux distribution that encompasses (1) the cell growth reaction and (2) the target metabolite production reaction, then the reactions not utilized by the flux distribution are eliminated. 
The MCS method identifies the smallest possible set of gene knockouts, with computations based on Farkas Lemma \cite{farkas}.
MCS method has demonstrated its capability to compute reaction deletions for growth-coupled production for most target metabolites in common model microorganisms like \textit{E. coli} and \textit{S. cerevisiae} \cite{von2017growth}. 
Furthermore, these methods have been successfully extended and applied to calculate the growth-coupled production of valuable metabolites, including itaconic acid and 2,3-butanediol in \textit{E. coli} \cite{schneider2020extended}, as well as indigoidine in \textit{P. putida} \cite{banerjee2020genome}. 
More recently, Tamura et al. introduced \gDel \cite{tamura2023gene}, a tool to calculate gene deletion strategies.
It operates by maximizing the number of repressed reactions, thereby extracting essential core components for growth-coupled production.
Experimental validations have demonstrated that in comparison with other methods, \gDel stands out as one of the most effective approaches currently available for calculating gene deletions. 
However, despite numerous efforts, these approaches exhibit a common shortcoming. 
All of the present methods employ de novo calculation strategies when deriving gene deletion strategies for various target metabolites.
This approach overlooks the shared information among different target metabolites, potentially leading to an excessive number of unnecessary repeated calculations.

On the other side, efforts have been made to establish databases on gene deletion strategies to address the growing demand for computational experiment data in strain design. 
A noteworthy example is the MetNetComp \cite{tamura2023metnetcomp} database. 
It is a web-based system that offers information on gene deletion strategies for growth-coupled production in constraint-based metabolic networks. 
MetNetComp computed growth-coupled gene deletion strategies that are minimal or maximal regarding the number of gene deletions by modifying the gDel\_minRN strategies.
It now houses a vast repository of over 85,000 gene-deletion strategies for metabolites across various constraint-based models from different species.
While the advent of such databases has significant potential to foster new research paradigms in the field, the exploration of their utility, especially concerning computational gene deletion strategies, is still in its infancy.

Therefore, it would be desirable to develop a method to combine existing database resources to alleviate the computational burden and efficiently calculate gene deletion strategies. 
However, extracting information that contributes to efficient computation from related databases and effectively integrating it into algorithms is not a straightforward task.

In this study, the authors propose \proposed, a novel database-enhance framework for calculating gene deletion strategies for growth-coupled production on genome-scale models. 
\proposed extracts information on gene deletion from the related database to boost the computational efficiency of downstream algorithms.
\proposed comprises two steps: (1) \textbf{STEP 1}, extracting the gene set from a gene deletion database that corresponds to the essential core components for growth-coupled production for a given constraint-based model;
and (2) \textbf{STEP 2}, using an extended version of \gDel that incorporates the gene set extracted from the STEP 1 as the initial gene pool.
The initial gene pool contains genes that are never knocked out in the gene deletion database, and these are protected from knockouts to narrow the algorithmic search space.

In the computational experiments, we compared \proposed with GDLS \cite{lun2009large}, optGene \cite{rocha2008optgene}, \gDel \cite{tamura2023gene}, and geneMCSEnumerator (gMCSE) \cite{von2017growth}.
Besides the aforementioned algorithm \gDel, gMCSE is an MCS-based algorithm for growth-coupled strain design using gene knockouts, which is available as an API function 'geneMCSEnumerator2' in the CellNetAnalyzer \cite{CellNetAnalyzer}, and we evaluated its first resulting mcs.
GDLS and optGene are also the most widely used software to derive gene deletion strategies, which are available in the COBRA Toolbox \cite{heirendt2019creation}.

All these methods were applied to three constraint-based models including e\_coli\_core \cite{orth2010reconstruction}, iMM904 \cite{mo2009connecting}, and iML1515 \cite{monk2017ml1515}.
e\_coli\_core is a constraint-based model that contains the only essential part of the metabolism of \textit{E. coli};
iMM904 and iML1515 are genome-scale constraint-based models of \textit{S. cerevisiae} and \textit{E. coli}, respectively; 

The average elapsed time of \proposed for e\_coli\_core, iMM904, and iML1515 were 1.12s, 79.92s, and 431.75s, respectively, which were substantially faster than all other compared methods.
In the meantime, \proposed achieves the success ratio for e\_coli\_core, iMM904, and iML1515 at 60.0\%, 11.5\%, and 51.2\%, which were substantially better than GDLS, optGene, and gMCSE, and closely similar to \gDel.

To the best of our knowledge, this study is the first attempt to extract information from pre-existing knowledge in gene deletion databases and integrate it with algorithms to derive new gene deletion strategies.
Furthermore, in comparison to existing approaches, our proposed method achieves a decent tradeoff between the success rate and time-consuming: it attains noteworthy enhancements in computational efficiency, all while maintaining a high success rate.

The remaining sections of this paper are organized as follows: 
Section \ref{sec:definition} formularizes several fundamental concepts in this study;
Section \ref{sec:problem} and \ref{sec:problem_example} describes the main problem of this study mathematically and illustrates it with a small example; 
Section \ref{sec:method} illustrates the proposed framework \proposed with a small example and provides the corresponding pseudo-code;
Section \ref{sec:experiments_setting} describes the basic experiment setting; 
Section \ref{sec:results} describes the experiment results: (1) the performance comparison of \proposed, GDLS, optGene, \gDel, and gMCSE for e\_coli\_core, iMM904, and iML1515, (2) the performance comparison of \proposed based on different initial gene pools, including Predicted-$G_{\text{remain}}$ genes as the default setting, $G_{\text{remain}}$ genes, growth essential genes, and randomly chosen genes;
Section \ref{sec:discussion} analyzes the results of the experiments, evaluates the performance of \proposed and other methods, and discusses future work.

\section{PRELIMINARY AND PROBLEM SETTING}
\label{sec:p_ps}
In this section, we first define several terms used in this study.
Then we introduce the main problem of this study and explain it with a toy constraint-based model.
All the notations mentioned in this section are listed in Table \ref{table: notations}.
\subsection{Definition}
\label{sec:definition}

\begin{table}[t]
  \caption{Notations used in the definitions in Section \ref{sec:p_ps}.}
  \centering
  \label{notation}
  \resizebox{0.95\linewidth}{!}{%
      \begin{tabular}{l|l}
          \toprule
          \textbf{Notation}    & \textbf{Description} \\
          \midrule
          \midrule
          $C$       & Constraint-based model\\
          $C_1$       & Metabolic network\\
          $C_2$       & GPR rule\\
          \midrule
          $M$       & Metabolites in the constraint-based model\\
          $R$       & Reactions in the constraint-based model\\
          $V$       & Reaction rates (flux)\\
          $S$       & Stoichiometry matrix\\
          $L$       & Lower bounds for reaction rates\\
          $U$       & Upper bounds for reaction rates\\
          \midrule
          $G$       & Genes in the constraint-based model\\
          $F$       & Boolean functions for GPR rule\\
          $P$       & Outputs of Boolean functions in $F$\\
          \midrule
          $m_{\text{target}}$       & Target metabolite\\
          $r_{\text{target}}$       & Target metabolite production reaction\\
          $r_{\text{growth}}$       & Cell growth reaction\\
          $PR_{\text{threshold}}$        & Lower bound of target production reaction rate\\
          $GR_{\text{threshold}}$        & Lower bound of cell growth reaction rate\\
          $D$       & Genes deleted in the gene deletion strategy\\
          \bottomrule
      \end{tabular}
      }
\label{table: notations}
\end{table}

\subsubsection{\textbf{Constraint-based model}}
Let $C = \{M, R, S, L, U, G, F, P\}$ be a constraint-based model.
The elements of $C$ can be further divided into two parts: (1) the metabolic network and (2) the GPR rule. 
We describe them as follows:

\begin{itemize}
  \item 
  Metabolic network: Let $C_1 = \{M, R, S, L, U\}$ be a metabolic network.
  $M = \{m_1, \ldots, m_a\}$ denotes a set consisting all metabolites, with one of them being the target metabolite $m_{\text{target}}$.
  $R = \{r_1, \ldots, r_b\}$ denotes a set consisting all reactions, including the cell growth reaction $r_{\text{growth}}$ and the target metabolite production reaction $r_{\text{target}}$.
  To facilitate clarity, we introduce an additional set $V = \{v_1, \ldots, v_b\}$ to represent the reaction rates per unit time (flux) corresponding to reactions in $R$. 
  Notably, we distinguish the rate of the growth reaction ($v_{\text{growth}}$) as the \textbf{Growth Rate (GR)} and that of the target metabolite production reaction ($v_{\text{target}}$) as the \textbf{Production Rate (PR)}.
  The stoichiometry matrix $S$ contains elements $S_{ij} = k$, indicating that reaction $r_j$ either produces (+) or consumes (-) $k$ units of metabolite $m_i$ per unit time.
  Lower and upper bounds for the reaction rates within $V$ are denoted by $L = \{l_1, \ldots, l_b\}$ and $U = \{u_1, \ldots, u_b\}$, respectively.

  \item
  GPR rule: Let $C_2 = \{G, F, P\}$ be a GPR rule.
  $G = \{g_1, \ldots, g_c\}$ represents a set of genes, while $F = \{f_1, \ldots, f_b\}$ represents Boolean functions.
  $P = \{p_1, \ldots, p_b\}$ is a collection of outputs generated by applying the functions in $F$ to the gene set $G$. 
  Each output $p_j$ can be expressed as $p_j = f_j(G)$, where both $p$ and $g$ are binary values, i.e., $p, g \in \{0,1\}$.
  Note that if $p_j = 0$, this imposes a constraint on both the lower bound $l_j$ and the upper bound $u_j$ to be 0, effectively restraining reaction $r_j$.
\end{itemize}

\subsubsection{\textbf{Flux balance analysis}}
When analyzing the metabolic network within a constraint-based model, flux balance analysis (FBA) assumes steady states where all metabolic reaction rates (fluxes) are constant \cite{orth2010flux}.
Specifically, we give the following definitions for FBA: (1) for each compound, the sum of the producing fluxes is equal to the sum of the consuming fluxes; (2) in each reaction, the fluxes of substrates and products must satisfy the ratio in the chemical reaction equation, and (3) the upper and lower bounds are given for each flux. 
In the standard procedure for FBA on the given metabolic network $C_1$, the objective is to maximize the cell growth reaction rate $v_{\text{growth}}$ using the following linear programming (LP):

\begin{algorithm}
  \caption*{LP formalization of FBA for a constraint-based model $C_1$.}
  \textbf{Given:} $C_1$ \\
  \textbf{Maximize:} $v_{\text{growth}}$ 
  \begin{algorithmic}[1]
    \STATE \textbf{Such that:}
    \STATE \hspace{1em}$\sum_{j}S_{ij}v_{j} = 0$ for all $i$;
    \STATE \hspace{1em}$l_j \leq v_j \leq u_j$ for all $j$;
    \STATE \hspace{1em}$i = \{1, \ldots, a\}$, $j = \{1, \ldots, b\}$
  \end{algorithmic}
  \label{algo: LP}
\end{algorithm}

\subsubsection{\textbf{Growth-coupled production}}
Growth-coupled production is a special situation of microorganism’s metabolic processes, wherein the cell growth co-occurs with the synthesis of the target metabolite.
During the simulation of constraint-based models, growth-coupled production can be defined by two indexes: the lower bound of target metabolite production rate ($PR_{\text{threshold}}$) and the lower bound of cell growth reaction rate ($GR_{\text{threshold}}$). 
When the GR is maximized, if the simulated values $v_{\text{target}}$ and $v_{\text{growth}}$ meet above two lower bounds in any cases, respectively, growth-coupled production is considered achieved. 
In this study, we set $PR_{\text{threshold}} = GR_{\text{threshold}} = 0.001$.

\subsection{Problem Definition}
\label{sec:problem}
Based on the above definitions, the primary objective of this study is to find the set $D$ consists of genes that need to be deleted to achieve the growth-coupled production of $m_{\text{target}}$, from which $r_{\text{target}}$ can easily be derived.
We define the main problem in a structured algorithmic form as follows:

\begin{algorithm}
  \caption*{Main problem formulation: identifying the deleted gene set $D$.}
  \textbf{Given:} $C, r_{\text{target}}, PR_{\text{threshold}}, GR_{\text{threshold}}$ \\
  \textbf{Find:} $D \subset G$ that results in $v_{\text{target}} \geq PR_{\text{threshold}}$, $v_{\text{growth}} \geq GR_{\text{threshold}}$ 
  \begin{algorithmic}[1]
    \STATE \textbf{Such that minimize:}\\
    \STATE \hspace{1em}$v_{\text{target}}$\\
    \STATE \hspace{1em}\textbf{Such that maximize:}\\
    \STATE \hspace{2em}$v_{\text{growth}}$\\
    \STATE \hspace{2em}\textbf{Such that:}\\
    \STATE \hspace{3em}$\sum_{j}S_{ij}v_{j} = 0$ for all $i$;\\
    \STATE \hspace{3em}
    $\begin{cases}
      v_j = 0 \text{ if } p_j = 0,\\
      l_j \leq v_j \leq u_j \text{, otherwise};
      \end{cases}$\\
      \STATE \hspace{3em}$p_j = f_j(G)$;\\
      \STATE \hspace{3em}$\begin{cases}
      g = 0 \text{ if } g \in D,\\
      g = 1 \text{, otherwise};
      \end{cases}$\\
  \end{algorithmic}
  \label{algo: problem}
\end{algorithm}

\subsection{Example}
\label{sec:problem_example}
Following is a small example with detailed explanations, to further illustrate the main problem addressed in this study. 
\begin{figure}[t]
  \centering
  \includegraphics[width=0.99\linewidth]{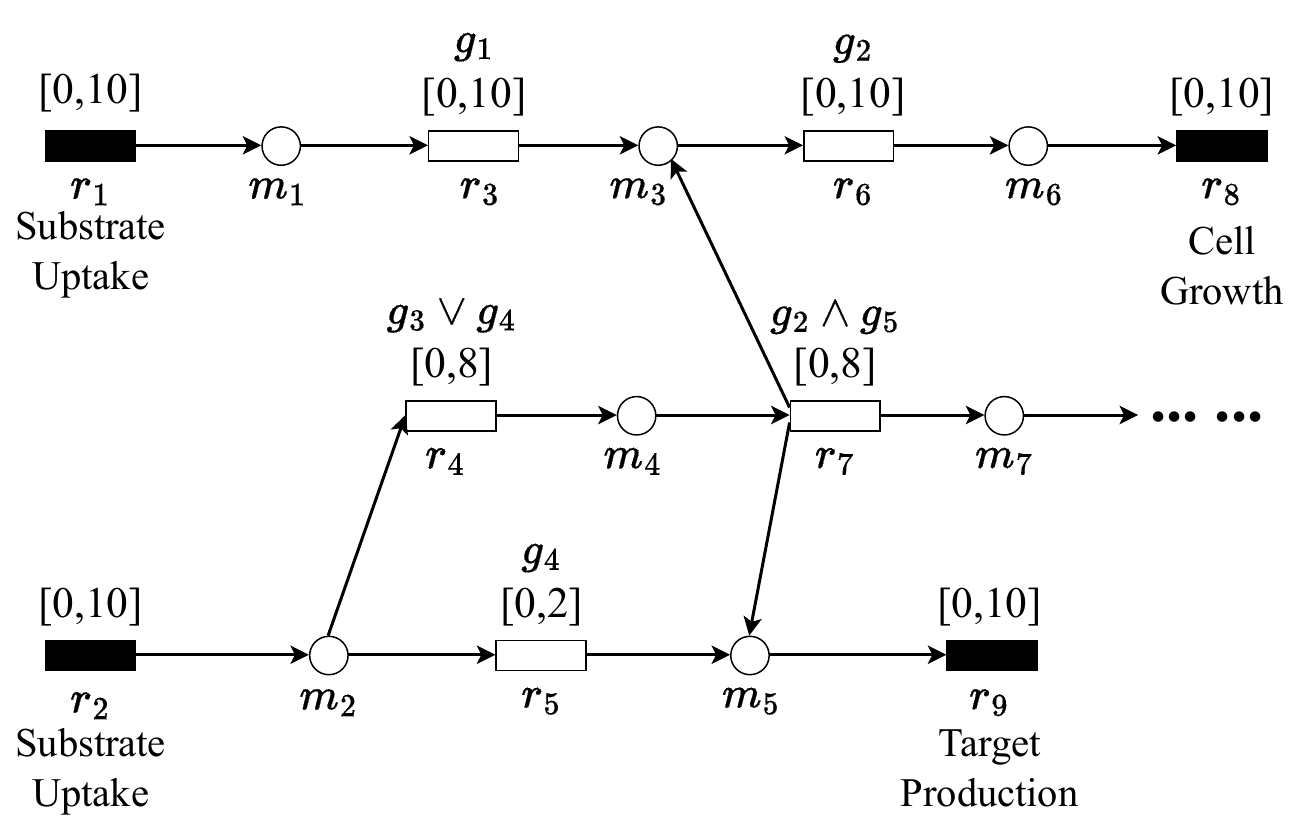}
  \caption{A toy example of the constraint-based model where circles and rectangles represent metabolites and reactions, respectively. 
  Black rectangles denote external and internal reactions. 
  $r_1$, $r_2$ correspond to two substrate uptake reactions.
  $r_8$, $r_{9}$ correspond to cell growth, and target metabolite production reactions, respectively. 
  The reaction rates are constrained by the range $[l_i, u_i]$. 
  This example shows only part of the model, the rest after $m_7$ is omitted.}  
  \label{fig:toy model}
\end{figure}

\subsubsection{\textbf{Toy constraint-based model settings}}
Fig. \ref{fig:toy model} shows a toy example and $C = (M, R, S, L, U, G, F, P)$ represents a constraint-based model.
Specifically, for its metabolic network part $C_1$, we have $M = \{m_1, \ldots, m_7\}$, $R = \{r_1, \ldots, r_{9}\}$, 
$S = \begin{pmatrix}
  1 & 0 & -1& 0 & 0 & 0 & 0 & 0 & 0 \\
  0 & 1 & 0 & -1& -1& 0 & 0 & 0 & 0 \\
  0 & 0 & 1 & 0 & 0 & -1& 1 & 0 & 0 \\
  0 & 0 & 0 & 1 & 0 & 0 & -1& 0 & 0 \\
  0 & 0 & 0 & 0 & 1 & 0 & 1 & 0 & 0 \\
  0 & 0 & 0 & 0 & 0 & 1 & 0 & -1& 0 \\
  0 & 0 & 0 & 0 & 0 & 0 & 1 & 0 & -1 \\
\end{pmatrix}$, $L = \{0, 0, 0, 0, 0, 0, 0, 0, 0\}$, and $U = \{10, 10, 10, 8, 2, 10, 8, 10, 10\}$.
For its gene reaction rule part $C_2$, we have $G = \{g_1, \ldots, g_5\}$, $F = \{f_1, \ldots, f_{9}\}$, and $P = \{p_1, \ldots, p_{9}\}$, where 
\begin{align*}
  f_i : p_i = 
  \begin{cases}
    1, &   i = 1\\ 
    1, &  i = 2\\ 
    g_1, &  i = 3\\ 
    g_3\vee g_4, &  i = 4\\ 
    g_4, &  i = 5\\ 
    g_2, &  i = 6\\ 
    g_2\wedge g_5, &  i = 7\\
    1, &  i = 8\\
    1, &  i = 9.
  \end{cases}
\end{align*}
 
In this toy model $C$, because the gene reaction rule for $r_7$ is given as $f_7: p_7 = g_2\wedge g_5$, the reaction rate of $r_7$ (denoted as $v_7$) is forced to be 0 if one of $g_2$ or $g_5$ is 0 (deleted), while $0 \leq v_7 \leq 8$ is held if both of $g_2$ and $g_5$ are 1 (not deleted).
However, for $r_4$, its reaction rate $v_4$ is forced to be 0 only if both $g_3$ and $g_4$ are 0, while $0 \leq v_4 \leq 8$ is held if at least one of $g_3$ or $g_4$ is 1.
For $r_1$, $r_2$, $r_8$ and $r_9$, since $p_1 = p_2 = p_8 = p_9 = 1$ always holds, none of $v_1$, $v_2$, $v_8$ or $v_9$ can be forced to be 0 by any gene deletions.

In this example, we set the $m_5$ as the target metabolite produced by the target metabolite production reaction $r_9$.
\begin{table}
  \centering
  \caption{Gene deletion strategies for the toy example in Fig. \ref{fig:toy model} are classified into eight types according to the resulting flux distributions (reaction rates).}
  \label{table: toy_1_sum}
  \resizebox{0.98\linewidth}{!}{%
  \begin{tabular}{cl} 
  \toprule
  Type & Gene deletion strategies classified by flux distributions \\ 
  \midrule
  1    & $\emptyset$, \{$g_3$\} \\
  2    & \{$g_1$\}, \{$g_1, g_3$\} \\
  3    & \{$g_2$\}, \{$g_1, g_2$\}, \{$g_1, g_5$\}, \{$g_2, g_3$\}, \{$g_2, g_5$\}, \{$g_1, g_2, g_3$\}, \\ 
       & \{$g_1, g_2, g_5$\}, \{$g_1, g_3, g_5$\}, \{$g_2, g_3, g_5$\}, \{$g_1, g_2, g_3, g_5$\}\\
  4    & \{$g_4$\} \\
  5    & \{$g_5$\}, \{$g_3, g_5$\} \\
  6    & \{$g_1, g_4$\} \\
  7    & \{$g_2, g_4$\}, \{$g_1, g_2, g_4$\}, \{$g_1, g_3, g_4$\}, \{$g_1, g_4, g_5$\}, \{$g_2, g_3, g_4$\}, \\ 
       & \{$g_2, g_4, g_5$\}, \{$g_1, g_2, g_3, g_4$\}, \{$g_1, g_2, g_4, g_5$\}, \{$g_1, g_3, g_4, g_5$\}, \\ 
       & \{$g_2, g_3, g_4, g_5$\}, \{$g_1, g_2, g_3, g_4, g_5$\}\\
  8    & \{$g_3, g_4$\}, \{$g_4, g_5$\}, \{$g_3, g_4, g_5$\} \\
  \bottomrule
  \end{tabular}
  }
\end{table}

\begin{table}
  \centering
  \caption{The resulting flux distribution of Types 1 to 8 gene deletion strategies.   For each gene deletion strategy type, the most optimistic (best) and pessimistic (worst) PR ($v_{9}$) at GR ($v_8$) maximization are shown.}
  \label{table: toy_1_fd}
  \resizebox{0.98\linewidth}{!}{%
  \begin{tabular}{cccccccccccc} 
  \toprule
  \multirow{2}{*}{ID} & \multirow{2}{*}{\begin{tabular}[c]{@{}c@{}}Gene deletion\\strategy\end{tabular}} & \multirow{2}{*}{\begin{tabular}[c]{@{}c@{}}PR\\situation\end{tabular}} & \multicolumn{9}{c}{Flux distribution}                                                \\ 
  \cmidrule{4-12}
                      &                                                                                   &                                                                        & $v_1$ & $v_2$ & $v_3$ & $v_4$ & $v_5$ & $v_6$ & $v_7$ & $v_8$               & $v_9$  \\ 
  \midrule
  1                   & \multirow{2}{*}{Type 1}                                                           & best                                                                   & 2     & 10    & 2     & 8     & 2     & 10    & 8     & \multirow{2}{*}{10} & 10     \\
  2                   &                                                                                   & worst                                                                  & 10    & 0     & 10    & 0     & 0     & 10    & 0     &                     & 0      \\ 
  \midrule
  3                   & \multirow{2}{*}{Type 2}                                                           & best                                                                   & 0     & 10    & 0     & 8     & 2     & 8     & 8     & \multirow{2}{*}{8}  & 10     \\
  4                   &                                                                                   & worst                                                                  & 0     & 8     & 0     & 8     & 0     & 8     & 8     &                     & 8      \\ 
  \midrule
  5                   & \multirow{2}{*}{Type 3}                                                           & best                                                                   & 0     & 2     & 0     & 0     & 2     & 0     & 0     & \multirow{2}{*}{0}  & 2      \\
  6                   &                                                                                   & worst                                                                  & 0     & 0     & 0     & 0     & 0     & 0     & 0     &                     & 0      \\ 
  \midrule
  7                   & \multirow{2}{*}{Type 4}                                                           & best                                                                   & 2     & 8     & 2     & 8     & 0     & 10    & 8     & \multirow{2}{*}{10} & 8      \\
  8                   &                                                                                   & worst                                                                  & 10    & 0     & 10    & 0     & 0     & 10    & 0     &                     & 0      \\ 
  \midrule
  9                   & \multirow{2}{*}{Type 5}                                                           & best                                                                   & 10    & 2     & 10    & 0     & 2     & 10    & 0     & \multirow{2}{*}{10} & 2      \\
  10                  &                                                                                   & worst                                                                  & 10    & 0     & 10    & 0     & 0     & 10    & 0     &                     & 0      \\ 
  \midrule
  11                  & \multirow{2}{*}{Type 6}                                                           & best                                                                   & 0     & 8     & 0     & 8     & 0     & 8     & 8     & \multirow{2}{*}{8}  & 8      \\
  12                  &                                                                                   & worst                                                                  & 0     & 8     & 0     & 8     & 0     & 8     & 8     &                     & 8      \\ 
  \midrule
  13                  & \multirow{2}{*}{Type 7}                                                           & best                                                                   & 0     & 0     & 0     & 0     & 0     & 0     & 0     & \multirow{2}{*}{0}  & 0      \\
  14                  &                                                                                   & worst                                                                  & 0     & 0     & 0     & 0     & 0     & 0     & 0     &                     & 0      \\ 
  \midrule
  15                  & \multirow{2}{*}{Type 8}                                                           & best                                                                   & 10    & 0     & 10    & 0     & 0     & 10    & 0     & \multirow{2}{*}{10} & 0      \\
  16                  &                                                                                   & worst                                                                  & 10    & 0     & 10    & 0     & 0     & 10    & 0     &                     & 0      \\
  \bottomrule
  \end{tabular}
  }
\end{table}

\subsubsection{\textbf{Gene deletions on the toy constraint-based model}}
Table \ref{table: toy_1_sum} describes the patterns of gene deletions in this example: $2^5 = 32$ patterns are classified into eight cases according to the flux distributions.
In the original state of the toy model, where no genes are deleted, when we maximize GR, we can obtain its maximum value $\max(GR) = \max(v_8) = 10$. 
However, there are two paths to reach from the substrate uptake reactions $r_1$ and $r_2$ to the growth reaction $r_8$: $(r_1 \rightarrow r_3 \rightarrow r_6 \rightarrow r_8)$ and $(r_2 \rightarrow r_4 \rightarrow r_7 \rightarrow r_6 \rightarrow r_8)$.
Besides, there is another path to reach from the substrate uptake reactions $r_2$ to the target production reaction $r_{9}$: $(r_2 \rightarrow r_5 \rightarrow r_{9})$.
If all three paths mentioned above are used, and the second and third paths achieve maximum fluxes, $GR=10$ and $PR=10$ can be obtained, as shown in ID 1 in Table \ref{table: toy_1_fd}. 
This is the most optimistic case regarding the value of PR. 
However, if only the first path is used, $GR=10$ and $PR=0$ can be obtained, as shown in ID 2 of Table \ref{table: toy_1_fd}. 
This is the most pessimistic case regarding the value of PR.
In this study, we evaluate the most pessimistic value of PR when the GR is maximized. 
Therefore, in the original state of the toy model, we have $GR=10$ and $PR=0$. 
Gene deletion strategy \{$g_3$\} is also classified as Type 1 because the same flux distribution is obtained.

\begin{figure*}[h]
  \centering
  \includegraphics[width=0.87\linewidth]{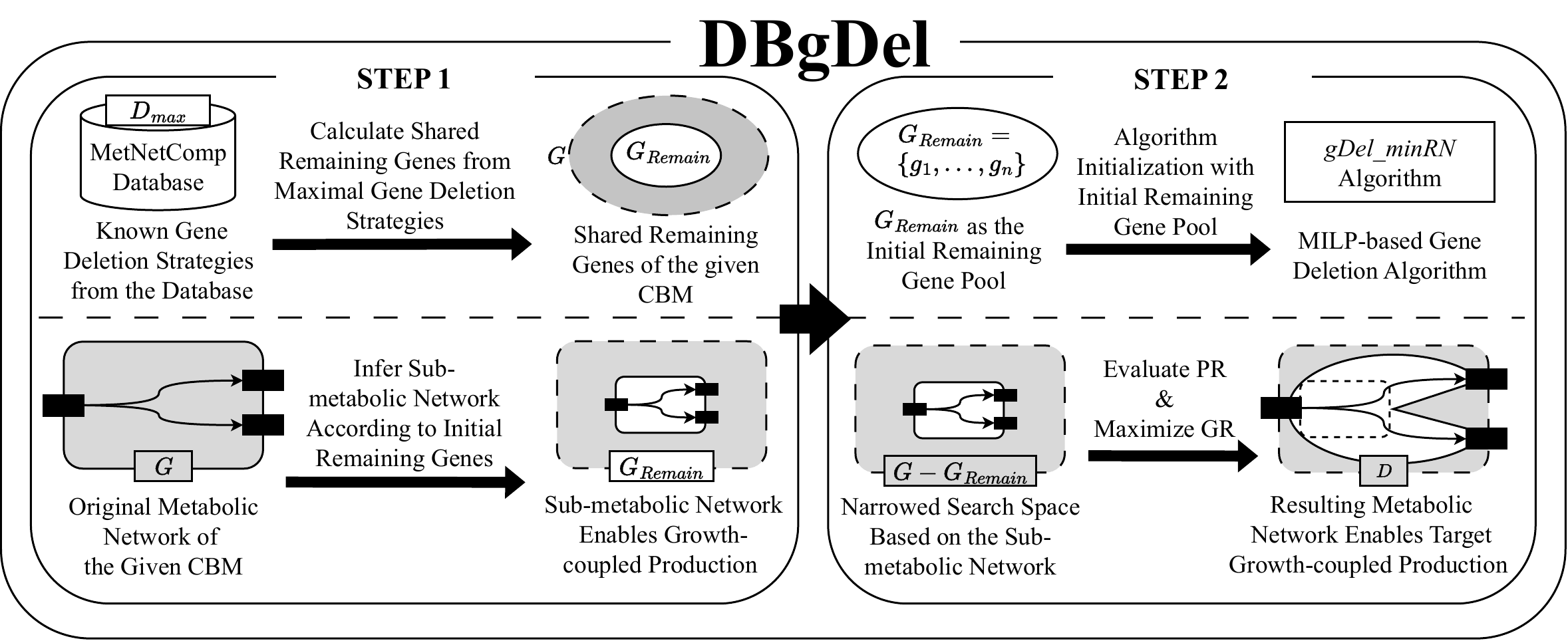}
  \caption{An overview of the proposed DBgDel framework. The \proposed framework comprises two steps: (1) \textbf{STEP 1}, \proposed takes known gene deletion strategies from the MetNetComp database as input and constructs a remaining gene set $G_{\text{remain}}$ as output; and (2) \textbf{STEP 2}, \proposed uses an extended version of \gDel algorithm that incorporates the $G_{\text{remain}}$ to calculate the deleted gene set $D$ for a new target metabolite, as the final output gene deletion strategies of the framework.}
  \label{fig:ov}
\end{figure*}

IDs 3 and 4 of Table \ref{table: toy_1_fd} describe the the most optimistic and pessimistic flux distributions regarding the PR value when $g_1$ is deleted.
When $g_1$ is deleted, $f_3: p_3 = g_1 = 0$ is obtained and $v_3$ is forced to be zero. 
The maximum GR is $v_8 = 8$ since only the second path reaches the growth reaction (from $r_2$ to $r_8$) can be used.
In the most optimistic case, the thrid path $(r_2 \rightarrow r_5 \rightarrow r_{9})$ that reach from the substrate uptake reaction $r_2$ to the target production reaction $r_{9}$ achieve maximum fluxes, therefore $GR=10$ and $PR=10$ are obtained. 
In the most pessimistic case, the above path is not used, therefore $GR=10$ and $PR=8$ are obtained. 
The gene deletion strategy \{$g_1, g_3$\} is also classified as Type 2 as shown in Table \ref{table: toy_1_fd}.

IDs 5 and 6 of Table \ref{table: toy_1_fd} describe the flux distributions when $g_2$ is deleted.
When $g_2$ is deleted, $f_6: p_6 = g_2 = 0$ and $f_7: p_7 = g_2 \wedge g_5 = 0$ are obtained, both $v_6$ and $v_7$ are forced to be zero. 
The maximum GR is $v_8 = 0$ since none of the two paths (from $r_1$ or $r_2$ to $r_8$) that reach cell growth reaction can be used.
In the most optimistic case, the path $(r_2 \rightarrow r_5 \rightarrow r_{9})$ that reach from the substrate uptake reaction $r_2$ to the target production reaction $r_{9}$ achieve maximum fluxes, therefore $GR=0$ and $PR=2$ are obtained. 
In the most pessimistic case, the above path is not used, therefore $GR=0$ and $PR=0$ are obtained. 
The gene deletion strategies \{$g_1, g_2$\}, \{$g_1, g_5$\}, \{$g_2, g_3$\}, \{$g_2, g_5$\}, \{$g_1, g_2, g_3$\}, \{$g_1, g_2, g_5$\}, \{$g_1, g_3, g_5$\}, \{$g_2, g_3, g_5$\}, and \{$g_1, g_2, g_3, g_5$\} are also classified as Type 3 as shown in Table \ref{table: toy_1_fd}.

IDs 7 and 8 of Table \ref{table: toy_1_fd} describe the flux distributions when $g_4$ is deleted.
When $g_4$ is deleted, $f_5: p_5 = g_4 = 0$ is obtained, and $v_5$ is forced to be zero. 
The maximum GR is $v_8 = 10$ since both of the two paths (from $r_1$ or $r_2$ to $r_8$) that reach cell growth reaction can be used.
In the most optimistic case, both two path above are used and the second one achieves maximum fluxes, therefore $GR=10$ and $PR=8$ are obtained. 
In the most pessimistic case, only the first path (from $r_1$ to $r_8$) is used, therefore $GR=10$ and $PR=0$ are obtained.

IDs 9 and 10 of Table \ref{table: toy_1_fd} describe the flux distributions when $g_5$ is deleted.
When $g_5$ is deleted, $f_7: p_7 = g_2 \wedge g_5 = 0$ is obtained, and $v_7$ is forced to be zero. 
The maximum GR is $v_8 = 10$ since only the first path (from $r_1$ to $r_8$) that reaches cell growth reaction can be used.
In the most optimistic case, the path $(r_2 \rightarrow r_5 \rightarrow r_{9})$ that reach from the substrate uptake reaction $r_2$ to the target production reaction $r_{9}$ is used and achieves maximum fluxes, therefore $GR=10$ and $PR=2$ are obtained. 
In the most pessimistic case, only the first path (from $r_1$ to $r_8$) is used, therefore $GR=10$ and $PR=0$ are obtained. 
The gene deletion strategy \{$g_3, g_5$\} is also classified as Type 5 as shown in Table \ref{table: toy_1_fd}.

IDs 11 and 12 of Table \ref{table: toy_1_fd} describe the flux distributions when \{$g_1, g_4$\} are deleted.
When \{$g_1, g_4$\} are deleted, $f_3: p_3 = g_1 = 0$ and $f_5: p_5 = g_4 = 0$ are obtained, both $v_3$ and $v_5$ are forced to be zero. 
The flux distribution is uniquely determined, $v_1 = 0$, $v_2 = v_4 = v_6 = v_7 = 8$ are obtained, and GR and PR are 8 for both optimistic and pessimistic cases regarding the PR value. 

IDs 13 and 14 of Table \ref{table: toy_1_fd} describe the flux distributions when \{$g_2, g_4$\} are deleted.
When \{$g_2, g_4$\} are deleted, $f_5: p_5 = g_4 = 0$ and $f_6: p_6 = g_2 = 0$ are obtained, both $v_5$ and $v_6$ are forced to be zero. 
The maximum GR is $v_8 = 0$ since none of the two paths (from $r_1$ or $r_2$ to $r_8$) that reach cell growth reaction can be used.
The flux distribution is uniquely determined where all reaction rates become zero, therefore PR is 0 for both optimistic and pessimistic cases regarding the PR value.
The gene deletion strategies \{$g_1, g_2, g_4$\}, \{$g_1, g_3, g_4$\}, \{$g_1, g_4, g_5$\}, \{$g_2, g_3, g_4$\}, \{$g_2, g_4, g_5$\}, \{$g_1, g_2, g_3, g_4$\}, \{$g_1, g_2, g_4, g_5$\}, \{$g_1, g_3, g_4, g_5$\}, \{$g_2, g_3, g_4, g_5$\}, and \{$g_1, g_2, g_3, g_4, g_5$\} are also classified as Type 7 as shown in Table \ref{table: toy_1_fd}.

IDs 15 and 16 of Table \ref{table: toy_1_fd} describe the flux distributions when \{$g_3, g_4$\} are deleted.
When \{$g_3, g_4$\} are deleted, $f_4: p_4 = g_3 \vee g_4 = 0$ and $f_5: p_5 = g_4 = 0$ are obtained, both $v_4$ and $v_5$ are forced to be zero. 
The maximum GR is $v_8 = 10$ since only the first path (from $r_1$ to $r_8$) can be used.
The flux distribution is uniquely determined, $v_1 = v_3 = v_6 = 10$, $v_2 = v_7 = 0$ are obtained, and PR is 0 for both optimistic and pessimistic cases regarding the PR value. 
The gene deletion strategies \{$g_4, g_5$\} and \{$g_3, g_4, g_5$\} are also classified as Type 8 as shown in Table \ref{table: toy_1_fd}.

Through all the cases above, we find that for gene deletion strategies of Types 2 and 6, even in the most pessimistic case, we still meet the criteria of $PR \geq PR_{\text{threshold}}$ and $GR \geq GR_{\text{threshold}}$. 
Consequently, in our example of the toy model, these two types of deletions enable growth-coupled production of the target metabolite $m_5$.

\section{Method}
\label{sec:method}
In this section, we first introduce the proposed \proposed framework.
Then we illustrate the behaviors of \proposed on the former constructed toy model with a new target metabolite, where an auxiliary reaction $r_{10}$ was added, as shown in Fig. \ref{fig:toy model_2}.

\subsection{\proposed Framework}
As shown in Fig. \ref{fig:ov}, The \proposed framework comprises two steps: (1) \textbf{STEP 1}, \proposed takes known gene deletion strategies from the MetNetComp database as input and constructs a remaining gene set $G_{\text{remain}}$ as output;
and (2) \textbf{STEP 2}, \proposed uses an extended version of \gDel algorithm that incorporates the $G_{\text{remain}}$ to calculate the deleted gene set $D$ for a new target metabolite, as the final output gene deletion strategies of the framework.
We introduce the details of each step below.

\subsubsection{\textbf{STEP 1}}
For a given constraint-based model, STEP 1 aims to extract the prior information about how the growth-coupled production of its metabolites is achieved.

The basic idea behind this aim is that for any given constraint-based model's metabolic network part \( C_1 \), there should be some core components, denoted as \( C_1^{Core} \), that are essential for achieving growth-coupled production for all of its metabolites (if possible). 

The \( C_1^{Core} \) is given by all the metabolites in a constraint-based model and forms a sub-network of the full metabolic network \( C_1 \).
Therefore, based on information about how various metabolites achieve growth-coupled production through sub-networks, we can make a prediction of \( C_1^{Core} \).

Such information can also be extracted from the gene deletion strategies, since each strategy gives the deleted genes $D$, according to which a sub-network of \( C_1 \) can be derived.

However, multiple gene deletion strategies may exist for a target metabolite to achieve its growth-coupled production. 
These strategies may result in gene deletions \( D \) of varying sizes, leading to redundant sub-networks of \( C_1 \) at different scales. 
To address this redundancy, we consider only the maximal gene deletion strategy \( D_{\text{max}}\) for each metabolite.
This strategy maximizes the size of the deleted gene set \( D \) to achieve growth-coupled production, corresponding to the smallest sub-network of \( C_1 \)
necessary for this purpose.

To achieve the above-illustrated aim of extracting prior information, in STEP 1, we construct a remaining gene set $G_{\text{remain}}$ based on the maximal gene deletion strategies \( D_{\text{max}}\) from the MetNetComp database.

Specifically, for a given constraint-based model, the $G_{\text{remain}}$ is given by:  
\begin{align}
  G_{\text{remain}} = G - (D_{\text{max}}^1 \cup D_{\text{max}}^2 \cup \ldots \cup D_{\text{max}}^{|M|})
\label{eq: gene_set}
\end{align}
where $G$ is a set of all genes, and \( |M| \) is the number of available maximal gene deletion strategies \( D_{\text{max}}\) for different metabolites in the MetNetComp database.

For a constraint-based model, when maximal gene deletion strategies are available for all its metabolites in the MetNetComp database, the resulting $G_{\text{remain}}$ exactly corresponds to $C_1^{\text{Core}}$ of this model. 

However, in most cases, when dealing with a new target metabolite without known gene deletion strategies for it, $G_{\text{remain}}$ is constructed based on information about the other metabolites. 
Therefore, as the default setting of STEP 1, we only track the maximal gene deletion strategies for a subset of the metabolites in the model, excluding the target itself. 
When predicting gene deletions for a new target metabolite based on information from other metabolites, we refer to the genes present in the resulting $G_{\text{remain}}$ as Predicted-$G_{\text{remain}}$ genes.

\subsubsection{\textbf{STEP 2}}
STEP 2 employs an extended version of the \gDel algorithm to calculate the gene set $D$ to be deleted. 
The \gDel is a mixed-integer linear programming (MILP)-based algorithm, which aims to identify growth-coupling gene deletion strategies that maximize the number of repressed reactions.

The MILP algorithm from Line 14 of the pseudocode searches for a candidate of the gene deletion strategy $D_{\text{candidate}}$ that satisfies the following criteria:
\begin{enumerate}
  \item Ensuring that the GR and PR are above the given thresholds $GR_{\text{threshold}}$ and $PR_{\text{threshold}}$, respectively.
  \item Maximizing the number of reactions repressed by gene deletions.
  \item Maximizing GR.
\end{enumerate}
Note that criterion (2) takes precedence over (3).
Then, starting from Line 26, $D_{\text{candidate}}$ is verified whether it achieves growth-coupled production of the target metabolite.
Here we incorporate the \gDel with the $G_{\text{remain}}$ from STEP 1 as the initial gene pool to narrow down the algorithmic search space.
As s result, the extended \gDel calculates gene deletion strategies without traversing the cases deleting genes in $G_{\text{remain}}$, and obtain the minimum reaction network for growth-coupled production.

The pseudocode of \proposed is given as follows.
\begin{algorithm}[H]
  \caption*{Procedure \proposed($C$, $v_{\text{target}}$, $maxloop$, $\alpha$, $\beta$)}
  \textbf{STEP1: Construct $G_{\text{remain}}$.}\\
  \begin{algorithmic}[1]
      \STATE $G_{\text{remain}} \gets G$ /*Initialize $G_{\text{remain}}$ as a copy of $G$*/
      \STATE $D_{\text{max\_union}} \gets \emptyset$ /*Initialize an empty set for the union of maximal gene deletion strategies*/
      \STATE /*Take $D_{\text{max}}^i$ for the $i$-th metabolite from database*/
      \FOR{i = 1 to $\left\lvert M \right\rvert$}
        \STATE $D_{\text{max\_union}} \gets D_{\text{max\_union}} \cup D_{\text{max}}^i$ 
      \ENDFOR
      \STATE $G_{\text{remain}} \gets G_{\text{remain}} - D_{\text{max\_union}}$ 
  \end{algorithmic}

  \vspace{0.5cm}

  \textbf{STEP2: Extended version of the \gDel algorithm.}\\
  \begin{algorithmic}[1]
    \STATE /* Calculate theoretical maximum PR and GR*/
      \STATE $PR_{\text{theormax}}$ = \textbf{max} $v_{\text{target}}$ \hfill /* theoretical maximum PR */
      \STATE \hspace{1em} \textbf{such that} $\sum_{j} S_{ij} v_{j} = 0$ for all $1 \leq i \leq a$
      \STATE \hspace{5.5em} $LB_j \leq v_j \leq UB_j$ for all $1 \leq j \leq b$
      \STATE $PR_{\text{threshold}} = \alpha \cdot PR_{\text{theormax}}$ 
      \STATE $GR_{\text{theormax}}$ = \textbf{max} $v_{\text{growth}}$ \hfill /* theoretical maximum GR */
      \STATE \hspace{1em} \textbf{such that} $\sum_{j} S_{ij} v_{j} = 0$ for all $1 \leq i \leq a$
      \STATE \hspace{5.5em} $LB_j \leq v_j \leq UB_j$ for all $1 \leq j \leq b$
      \STATE $GR_{\text{threshold}} = \beta \cdot GR_{\text{theormax}}$
      \STATE /* Finding a gene deletion strategy candidate.*/
      \STATE $prohibited\_list = \emptyset$, $loop = 1$
      \WHILE{$loop \leq maxloop$}
          \STATE /* maximize the number of repressed reactions first, GR second */
          \STATE \textbf{max} $GR_{\text{theormax}} \cdot KO + v_{\text{growth}}$
          \STATE /*KO: the number of repressed reactions.*/
          \STATE \textbf{such that} $\sum_{j} S_{ij} v_{j} = 0$ for all $1 \leq i \leq a$
          \STATE \hspace{1em} 
          $\begin{cases}
            v_j = 0 &\text{ if } p_j = 0,\\
            l_j \leq v_j \leq u_j, &\text{otherwise};
          \end{cases}$
          \STATE \hspace{1em} $p_j = f_j(G)$ 
          \STATE \hspace{1em} /* linear constraints for Boolean functions. */
          \STATE \hspace{1em} $g = 
          \begin{cases}
            0, &\text{ if } g \in D, \\
            1, &\text{ if } g \in G_{\text{remain}},\\
            1, &\text{otherwise};
          \end{cases}$
          \STATE \hspace{1em}/*D is the set of deleted genes.*/
          \STATE \hspace{1em}/*$G_{\text{remain}}$ as the initial remaining gene pool.*/

          \STATE \hspace{1em} $D \cap G_{\text{remain}} = \emptyset$ 
          \STATE \hspace{1em} $D \cap prohibited\_list = \emptyset$
          \STATE \hspace{1em} $ GR_{\text{threshold}} \leq v_{\text{growth}}$ and $ PR_{\text{threshold}} \leq v_{\text{target}}$
          \STATE $D_{\text{candidate}} = D$
    \STATE /*Check whether growth-coupled production is achieved.*/ 
    \STATE \textbf{min} $v_{\text{target}}$
    \STATE \hspace{1em} \textbf{such that max} $v_{\text{growth}}$
    \STATE \hspace{2em} \textbf{such that} 
    \STATE \hspace{3em} $\sum_{j} S_{ij} v_{j} = 0$ for all $i$
    \STATE \hspace{3em}
    $\begin{cases}
     v_j = 0 &\text{ if } p_j =0,\\
     l_j \leq v_j \leq u_j, &\text{otherwise};
    \end{cases}$
    \STATE \hspace{3em} $p_j =f_j(G)$
    \STATE \hspace{3em} 
    $g = \begin{cases}
      0 & \text{if } g \in D_{\text{candidate}}, \\
      1, & \text{otherwise};
    \end{cases}$
    \IF{$v_{\text{target}} \geq PR_{\text{threshold}}$ and $v_{\text{growth}} \geq GR_{\text{threshold}}$}
      \RETURN $D_{\text{candidate}}$, $v_{\text{target}}$, $v_{\text{growth}}$
    \ELSE
      \STATE $prohibited\_list = prohibited\_list \cup D_{\text{candidate}}$
      \STATE $loop = loop + 1$
    \ENDIF
      \ENDWHILE
  \end{algorithmic}
\end{algorithm}

\subsection{Example}
We apply the \proposed framework to the former constructed toy model with a new target metabolite to illustrate its execution.

\begin{figure}[t]
  \centering
  \includegraphics[width=0.99\linewidth]{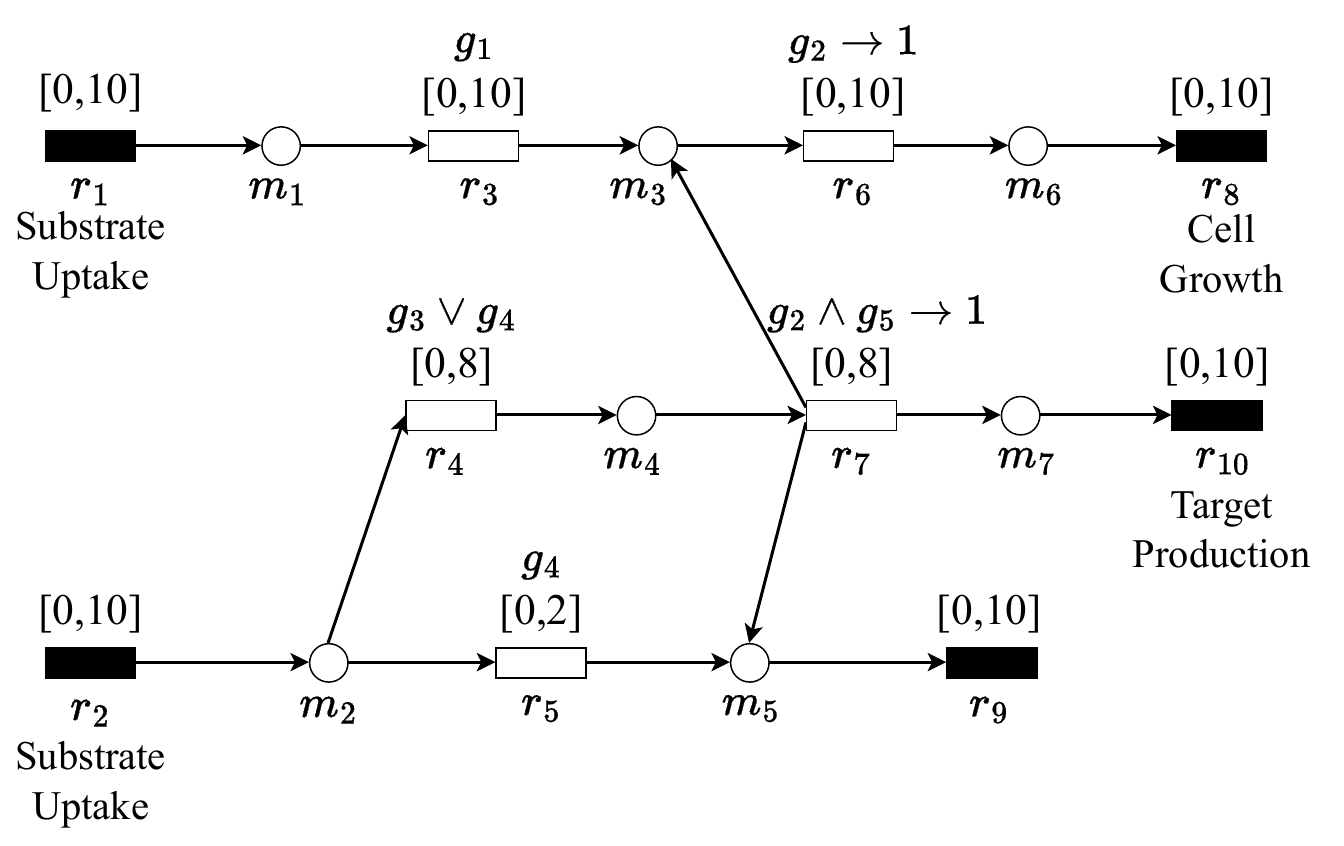}
  \caption{The toy model with a new target metabolite.
  Circles and rectangles represent metabolites and reactions, respectively. 
  Black rectangles denote external and internal reactions. 
  $r_1$, $r_2$ correspond to two substrate uptake reactions.
  $r_8$, $r_{10}$ correspond to cell growth, and target metabolite production reactions, respectively.
  The reaction rates are constrained by the range $[l_i, u_i]$.   
  }
  \label{fig:toy model_2}
\end{figure}

\subsubsection{\textbf{Toy constraint-based model settings}}
As shown in Fig. \ref{fig:toy model_2}, we set $m_7$ as the new target metabolite.
In the original model, $m_7$ does not have an external reaction, therefore it is necessary to add an auxiliary reaction $r_{10}$ to simulate its secretion.
The other comments of $C_1$ and $C_2$ part of the model are the same as the original one in Section. \ref{sec:p_ps}.

\begin{table}
  \centering
  \caption{The remaining genes correspond to the gene deletion strategies that enable growth-coupled production of the target metabolite in the toy example in Fig. \ref{fig:toy model}.}
  \begin{tabular}{cc} 
  \toprule
  Gene deletion strategies & Corresponding remaining genes  \\ 
  \midrule
  \{$g_1$\}           & \{$g_2, g_3, g_4, g_5$\}       \\
  \{$g_1, g_3$\}      & \{$g_2, g_4, g_5$\}            \\
  \{$g_1, g_4$\}      & \{$g_2, g_3, g_5$\}            \\
  \bottomrule
  \end{tabular}
  \label{table: remaining_genes}
\end{table}

\begin{table}
  \centering
  \caption{Gene deletion strategies for the toy example in Fig. \ref{fig:toy model_2} are classified into four types according to the resulting flux distributions (reaction rates).}
  \resizebox{0.98\linewidth}{!}{%
  \begin{tabular}{cl} 
  \toprule
  Type & Gene deletion strategies classified by flux distributions \\ 
  \midrule
  1    & $\emptyset$, \{$g_3$\}, \{$g_4$\} \\
  2    & \{$g_1$\}, \{$g_1, g_3$\}, \{$g_1, g_4$\} \\
  3    & \{$g_3, g_4$\} \\
  4    & \{$g_1, g_3, g_4$\} \\
  \bottomrule
  \end{tabular}
  }
  \label{table: toy_2_sum}
\end{table}

\begin{table}
  \centering
  \caption{The resulting flux distribution of Types 1 to 4 gene deletion strategies.   For each gene deletion strategy type, the most optimistic (best) and pessimistic (worst) PR ($v_9$) at GR ($v_8$) maximization are shown.}
  \label{table: toy_2_fd}
  \resizebox{0.98\linewidth}{!}{%
  \begin{tabular}{ccccccccccccc} 
  \toprule
  \multirow{2}{*}{ID} & \multirow{2}{*}{\begin{tabular}[c]{@{}c@{}}Gene deletion\\strategy\end{tabular}} & \multirow{2}{*}{\begin{tabular}[c]{@{}c@{}}PR\\situation\end{tabular}} & \multicolumn{10}{c}{Flux distribution}                                                      \\ 
  \cmidrule{4-13}
                      &                                                                                   &                                                                        & $v_1$ & $v_2$ & $v_3$ & $v_4$ & $v_5$ & $v_6$ & $v_7$ & $v_8$               & $v_9$ & $v_{10}$  \\ 
  \midrule
  1                   & \multirow{2}{*}{Type 1}                                                           & best                                                                   & 2     & 8     & 2     & 8     & 0     & 10    & 8     & \multirow{2}{*}{10} & 8     & 8     \\
  2                   &                                                                                   & worst                                                                  & 10    & 0     & 10    & 0     & 0     & 10    & 0     &                     & 0     & 0     \\ 
  \midrule
  3                   & \multirow{2}{*}{Type 2}                                                           & best                                                                   & 0     & 8     & 0     & 8     & 0     & 8     & 8     & \multirow{2}{*}{8}  & 8     & 8     \\
  4                   &                                                                                   & worst                                                                  & 0     & 8     & 0     & 8     & 0     & 8     & 8     &                     & 8     & 8     \\ 
  \midrule
  5                   & \multirow{2}{*}{Type 3}                                                           & best                                                                   & 10    & 0     & 10    & 0     & 0     & 10    & 0     & \multirow{2}{*}{10} & 0     & 0     \\
  6                   &                                                                                   & worst                                                                  & 10    & 0     & 10    & 0     & 0     & 10    & 0     &                     & 0     & 0     \\ 
  \midrule
  7                   & \multirow{2}{*}{Type 4}                                                           & best                                                                   & 0     & 0     & 0     & 0     & 0     & 0     & 0     & \multirow{2}{*}{0}  & 0     & 0     \\
  8                   &                                                                                   & worst                                                                  & 0     & 0     & 0     & 0     & 0     & 0     & 0     &                     & 0     & 0     \\
  \bottomrule
  \end{tabular}
  }
  \end{table}

\subsubsection{\textbf{Behaviors of the \proposed framework}}
In STEP 1, the \proposed framework constructs a gene set $G_{\text{remain}}$ for the initial remaining gene pool.
There are three known gene deletion strategies, i.e., $\{g_1\}$, $\{g_1, g_3\}$, and $\{g_1, g_4\}$ enable growth-coupled production of target metabolte in Fig.\ref{fig:toy model}. 
These strategies are used as inputs of STEP 1 to generate their corresponding remaining genes.
As shown in Table \ref{table: remaining_genes}, all these strategies correspond to remaining genes $g_2$ and $g_5$.
Consequently, we have Predicted-$G_{\text{remain}} = \{g_2, g_5\}$ for the target metabolite $m_7$.
For comparison, we also constructed the toy model's growth-essential gene set, denoted as $G_{GE}$, by examining its cell growth reaction situation. 
Specifically, since deleting $g_2$ always leads to $GR = 0$, in this example, we have $G_{GE} = \{g_2\}$.

Next in STEP 2, we set the initial remaining gene pool as Predicted-$G_{\text{remain}}$.
The GPR rules related to the genes in $G_{\text{remain}}$ are modified.
Specifically, for $g_2$ and $g_5$, we modify $f_6$ into $f_6: p_6 = g_2 = 1$, and $f_7$ into  $f_7: p_7 = g_2 \wedge g_5 = 1$.
According to the modified GPR rules, genes $g_2$ and $g_5$ will never be deleted, the gene deletion algorithm \gDel in STEP 2 doesn't need to compute their related gene deletions, leaving only three genes ($g_1$, $g_3$ and $g_4$) in the searching space.
Therefore, only $2^3 = 8$ patterns of gene deletions are left in this example, which are classified into four cases according to the flux distributions, as described in Table \ref{table: toy_2_sum}.
Similarly, in the case of using $G_{GE}$ as the initial remaining gene pool ($G_{\text{remain}} = G_{GE}$), the \gDel algorithm needs to compute $2^4 = 16$ patterns of gene deletions.

In the original state of the toy model, where no genes are deleted, when we maximize GR, we can obtain its maximum value $\max(GR) = \max(v_8) = 10$. 
However, there are two paths to reach from the substrate uptake reactions $r_1$ and $r_2$ to the growth reaction $r_8$: $(r_1 \rightarrow r_3 \rightarrow r_6 \rightarrow r_8)$ and $(r_2 \rightarrow r_4 \rightarrow r_7 \rightarrow r_6 \rightarrow r_8)$.
If both two paths mentioned above are used, and the second one achieves maximum fluxes, $GR=10$ and $PR=8$ can be obtained, as shown in ID 1 in Table \ref{table: toy_2_fd}. 
This is the most optimistic case regarding the value of PR. 
However, if only the first path is used, $GR=10$ and $PR=0$ can be obtained, as shown in ID 2 of Table \ref{table: toy_2_fd}. 
This is the most pessimistic case regarding the value of PR.
Therefore, in the original state of the toy model, we have $GR=10$ and $PR=0$. 
Gene deletion strategies \{$g_3$\} and \{$g_4$\} are also classified as Type 1 because the same flux distribution is obtained.

IDs 3 and 4 of Table \ref{table: toy_2_fd} describe the the most optimistic and pessimistic flux distributions regarding the PR value when $g_1$ is deleted.
When $g_1$ is deleted, $f_3: p_3 = g_1 = 0$ is obtained and $v_3$ is forced to be zero. 
The maximum GR is $v_8 = 8$ since only the second path (from $r_2$ to $r_8$) that reaches cell growth reaction can be used.
The flux distribution is uniquely determined, $v_1 = v_5 = 0$, $v_2 = v_4 = v_7 = v_9 = 8$ are obtained, and PR is 8 for both optimistic and pessimistic cases regarding the PR value. 
The gene deletion strategies \{$g_1, g_3$\} and \{$g_1, g_4$\} are also classified as Type 2 as shown in Table \ref{table: toy_2_fd}.

IDs 5 and 6 of Table \ref{table: toy_2_fd} describe the flux distributions when \{$g_3, g_4$\} are deleted.
When \{$g_3, g_4$\} are deleted, $f_4: p_4 = g_3 \vee g_4 = 0$ and $f_5: p_5 = g_4 = 0$ are obtained, both $v_4$ and $v_5$ are forced to be zero. 
The maximum GR is $v_8 = 10$ since only the first path (from $r_1$ to $r_8$) that reaches cell growth reaction can be used.
The flux distribution is uniquely determined, $v_1 = v_3 = v_6 = 10$, $v_2 = v_7 = v_9 = 0$ are obtained, and PR is 0 for both optimistic and pessimistic cases regarding the PR value. 

\begin{table*}
  \centering
  \caption{The constraint-based models that were used in the computational experiments.}
  \label{table3 }
  \begin{tabular}{lccc} 
  \toprule
  Model                                          & e\_coli\_core & iMM904 & iML1515  \\ 
  \midrule
  \#Genes                                        & 137           & 905    & 1516     \\
  \#Growth essential genes                       & 7             & 110    & 196      \\
  \#$G_{\text{remain}}$ genes                                 & 11            & 178    & 254      \\
  \#Predicted-$G_{\text{remain}}$ genes                              & 23            & 195    & 267      \\
  \#Randomly chosen genes                              & 23            & 195    & 267      \\  
  \midrule
  \#Reactions                                    & 95            & 1577   & 2712     \\ 
  \midrule
  \#Metabolites                                  & 72            & 1226   & 1877     \\
  \#Target metabolites                           & 48            & 782    & 1085     \\
  \#Target metabolites (extracellular space)     & 8             & 82     & 94       \\
  \#Target metabolites (non-extracellular space) & 40            & 700    & 991      \\
  \bottomrule
  \end{tabular}
  \label{table: model_sum}
\end{table*}

IDs 7 and 8 of Table \ref{table: toy_2_fd} describe the flux distributions when \{$g_1, g_3, g_4$\} are deleted.
When \{$g_1, g_3, g_4$\} are deleted, $f_3: p_3 = g_1 = 0$, $f_4: p_4 = g_3 \vee g_4 = 0$, and $f_5: p_5 = g_4 = 0$ are obtained, $v_3$, $v_4$ and $v_5$ are forced to be zero. 
The maximum GR is $v_8 = 0$ since none of the two paths (from $r_1$ or $r_2$ to $r_8$) that reach cell growth reaction can be used.
The flux distribution is uniquely determined where all reaction rates become zero, therefore PR is 0 for both optimistic and pessimistic cases regarding the PR value. 

Through all the cases above, we find that for gene deletion strategies of Type 2, even in the most pessimistic case, we still meet the criteria of $PR \geq PR_{\text{threshold}}$ and $GR \geq GR_{\text{threshold}}$. 
Consequently, in our example of the toy model, this type of deletions enable growth-coupled production of the target metabolite $m_7$.

\section{Computational Experiments}
\label{sec:computational experiments}

\subsection{Experiments Setting}
\label{sec:experiments_setting}

In the computational experiments, we applied the \proposed framework to three metabolic models \textit{e\_coli\_core}, \textit{iMM904}, and \textit{iML1515}.
The number of reactions, genes, and metabolites for the three metabolic models are summarized in Table \ref{table: model_sum}.

We define four types of genes used in the experiments, i.e., Growth Essential (GE) genes, $G_{\text{remain}}$ genes, Predicted-$G_{\text{remain}}$ genes, and Randomly Chosen (RC) genes as below:
\begin{itemize}
  \item 
  Growth Essential (GE) genes: A gene is defined as a growth essential gene when its deletion results in a maximum GR of zero in FBA analysis.
  \item 
  $G_{\text{remain}}$ genes: For a given metabolic model, we track the maximal gene deletion strategies for all its metabolites archived in the MetNetComp database, as outlined in Eq. \ref{eq: gene_set} in STEP 1. 
  Specifically, genes present in the resulting $G_{\text{remain}}$ are remained in the maximal gene deletion strategies for all the metabolites. 
  We defined them as $G_{\text{remain}}$ genes. 
  \item 
  Predicted-$G_{\text{remain}}$ genes: For a given metabolic model, we track only the maximal gene deletion strategies for the metabolites that exist in the extracellular space (metabolite whose name ends with '\_e') archived in the MetNetComp database.
  The calculation is outlined in Eq. \ref{eq: gene_set} in STEP 1.
  In this case, the resulting gene set is calculated based only on information about the extracellular space metabolites. 
  We defined them as Predicted-$G_{\text{remain}}$ gene.
  \item
  Randomly Chosen (RC) genes: A set of genes randomly chosen from the given metabolic model. 
  RC genes have no biological significance and are used to ensure a fair comparison of effectiveness in narrowing the search space. 
  The number of RC genes equals that of Predicted-$G_{\text{remain}}$ genes.
\end{itemize}

We design the computational experiments as follows:
\begin{enumerate} 
  \item Evaluate the overall computational efficiency of different gene deletion calculation methods.
  We compared the performance of \proposed with other baseline methods, including GDLS, optGene, \gDel, and gMCSE. 
  \item Evaluate the efficacy of different initial gene pool settings of \proposed.
  We compared the performance of \proposed with different initial gene pools in STEP2: Predicted-$G_{\text{remain}}$ genes, GE genes, $G_{\text{remain}}$ genes, and RC genes.
  Here, Predicted-$G_{\text{remain}}$ is the default choice of the initial gene pool, and $G_{\text{remain}}$ serves as the performance benchmark.
\end{enumerate}

Note that some metabolites cannot be produced through simulation-based growth-coupled production.
We determined and excluded such metabolites from the target metabolites by calculating their theoretical maximum PR.
When a target metabolite does not have an external (exchange) reaction, an auxiliary exchange reaction was temporarily added to the model to simulate the secretion. 
The unit of every reaction rate is mmol/gDW/h, which will be omitted hereafter for simplicity of notation.

All procedures in the computational experiments were implemented on a Ubuntu 20.04.6 LTS machine with an Intel Xeon Gold Processor with 2.30 GHz 64 cores/128 threads, 128 GB memory, and 1TB SSD. 
This workstation had CPLEX 12.10, COBRA Toolbox v3.0, CellNetAnalyzer ver.2023.1, and MATLAB R2019b installed and used for these analyses. 

\subsection{Results and Performance Comparison}
\label{sec:results}

\textbf{For \textit{e\_coli\_core}:}
Table \ref{table: e1_compare} summarizes the performance comparison between the proposed method \proposed (using Predicted-$G_{\text{remain}}$ genes as the initial remaining gene pool) with GDLS, optGene, \gDel, and gMCSE in the context of targeting 40 non-extracellular space metabolites of \textit{e\_coli\_core}.
Regarding the success rate, \proposed achieved success in 24 out of 40 target metabolites, meeting the minimum GR and PR threshold of 0.001 or higher during GR maximization. 
In comparison, GDLS succeeded in only 5 out of 40 cases, optGene in 22 out of 40 cases, \gDel in 28 out of 40 cases, and gMCSE in 7 out of 40 cases.
Regarding computational efficiency, optGene required significantly more extended time, with an average of 1779.30 seconds, while the other methods cost about 1 second.

Table \ref{table: e1_gene} summarizes the performance comparison between the proposed method \proposed with three different initial remaining gene pool settings.  
When utilizing Predicted-$G_{\text{remain}}$ genes as the initial remaining gene pool, the success rate in achieving the minimum GR and PR threshold of 0.001 or higher during GR maximization was 24 out of 40 instances. 
Similarly, when using the GE genes as the initial remaining gene pool, the success rate was 28 out of 40 instances. 
When employing $G_{\text{remain}}$ genes as the initial remaining gene pool, the success rate remained at 28 out of 40 instances.
Additionally, when employing RC genes as the initial remaining gene pool, the success rate decreased to 21 out of 40 instances.
As for computational efficiency, since the model scale of \textit{e\_coli\_core} is relatively small, all four settings of the initial remaining gene pool can achieve a computation time of around 1 second, which is short enough.

\begin{table}
  \centering
  \caption{The performance comparison between the proposed method \proposed (using Predicted-$G_{\text{remain}}$ genes as the initial remaining gene pool) with GDLS, optGene, \gDel, and gMCSE in the situation of targeting non-extracellular space metabolites of \textit{e\_coli\_core}. 
  Each gene deletion strategy was deemed successful if it achieved a minimum GR and PR of 0.001 or higher during GR maximization. The unit of elapsed time is second (s).}
  \label{table: e1_compare}
  \resizebox{0.99\linewidth}{!}{%
  \begin{tabular}{lccccc} 
  \toprule
  Method           & \begin{tabular}[c]{@{}c@{}}Proposed\\framework\end{tabular} & GDLS  & gMCSE  & optGene & gDel\_minRN  \\ 
  \midrule
  Success rate     & 24/40  & 5/40 &   7/40  & 22/40   & 28/40     \\
  Avg. time & 1.12  & 1.24 &   1.73 & 1779.30 & 1.62   \\
  \bottomrule
  \end{tabular}
  }
\end{table}

\begin{table}
  \centering
  \caption{The performance comparison between the proposed method \proposed with three different initial remaining gene pool settings for \textit{e\_coli\_core}. 
  Each gene deletion strategy was deemed successful if it achieved a minimum GR and PR of 0.001 or higher during GR maximization. The unit of elapsed time is second (s).}
  \label{table: e1_gene}
  \resizebox{0.99\linewidth}{!}{%
  \begin{tabular}{lcccc} 
  \toprule
  \begin{tabular}[c]{@{}l@{}}Initial remaining \\gene pool\end{tabular} & \begin{tabular}[c]{@{}c@{}}Predicted-$G_{\text{remain}}$ \\(Proposed)\end{tabular} & GE gene & $G_{\text{remain}}$  & RC gene \\ 
  \midrule
  Success rate                                                          & 24/40                                                               & 28/40                 & 28/40        & 21/40 \\ 
  \midrule
  Time.Success = 1                                                      & 0.59                                                              & 0.68                & 0.60      &  0.59\\
  Time.Success = 0                                                      & 1.91                                                              & 2.83                & 2.57      &  2.34\\
  Time.All cases                                                        & 1.12                                                              & 1.33                & 1.12     &  1.17\\
  \bottomrule
  \end{tabular}
  }
\end{table}

\textbf{For \textit{iMM904}:}
Table \ref{table: e2_compare} summarizes the performance comparison between the proposed method \proposed (using Predicted-$G_{\text{remain}}$ genes as the initial remaining gene pool) with GDLS, optGene, \gDel, and gMCSE in the context of targeting 700 non-extracellular space metabolites of \textit{iMM904}.
Regarding the success rate, \proposed achieved success in 81 out of 700 target metabolites, meeting the minimum GR and PR threshold of 0.001 or higher during GR maximization. 
In comparison, GDLS failed in all 700 cases, optGene succeeded in only 27 out of 700 cases, \gDel in 116 out of 700 cases, and gMCSE in 17 out of 700 cases.
\proposed demonstrated outstanding performance in average elapsed time, with a mean time of 79.92 seconds. 
GDLS had an average time of 243.42 seconds, optGene required significantly more extended time with an average of 1986.45 seconds, \gDel had an average time of 422.32 seconds, and gMCSE had an average time of 584.24 seconds.

Table \ref{table: e2_gene} summarizes the performance comparison between the proposed method \proposed with three different initial remaining gene pool settings. 
The search space of gDel\_minRN is larger than that of DBgDel, but the success rates of GE genes and $G_{remain}$ are higher.
This is because GE genes and $G_{remain}$ effectively utilize information about the genes that must not be deleted, which gDel\_minRN does not refer to.
When utilizing Predicted-$G_{\text{remain}}$ genes as the initial remaining gene pool, the success rate in achieving the minimum GR and PR threshold of 0.001 or higher during GR maximization was 81 out of 700 instances. 
When using the GE genes as the initial remaining gene pool, the success rate was 120 out of 700 instances. 
When employing $G_{\text{remain}}$ genes as the initial remaining gene pool, the success rate remained at 126 out of 700 instances.
Additionally, when employing RC genes as the initial remaining gene pool, the success rate decreased to 65 out of 700 instances.
Examining the computational efficiency, the proposed method demonstrated varied performance regarding elapsed time for successful cases. 
For instances where success was achieved, the average elapsed time was 22.89 seconds with Predicted-$G_{\text{remain}}$ genes, 189.90 seconds with GE genes, 47.82 seconds with $G_{\text{remain}}$ genes, and 31.93 seconds with RC genes.
In contrast, for cases where success was not achieved, the average elapsed time was 87.38 seconds with Predicted-$G_{\text{remain}}$ genes, 459.31 seconds with GE genes, 224.19 seconds with $G_{\text{remain}}$ genes, and 157.42 seconds with RC genes.
Considering all cases, regardless of success or failure, the proposed method showed outstanding average elapsed times when using Predicted-$G_{\text{remain}}$ genes, with values of 79.92 seconds.
The average elapsed times of 145.77 and 192.44 seconds remain competitive when using RC genes or $G_{\text{remain}}$ genes.

\begin{table}
  \centering
  \caption{The performance comparison between the proposed method \proposed (using Predicted-$G_{\text{remain}}$ genes as the initial remaining gene pool) with GDLS, optGene, \gDel, and gMCSE in the situation of targeting non-extracellular space metabolites of \textit{iMM904}. 
  Each gene deletion strategy was deemed successful if it achieved a minimum GR and PR of 0.001 or higher during GR maximization. The unit of elapsed time is second (s).}
  \label{table: e2_compare}
  \resizebox{0.99\linewidth}{!}{%
  \begin{tabular}{lccccc} 
  \toprule
  Method           & \begin{tabular}[c]{@{}c@{}}Proposed\\framework\end{tabular} & GDLS & gMCSE  & optGene & gDel\_minRN  \\ 
  \midrule
  Success rate     & 81/700    & 0/700  &    17/700   & 27/700   & 116/700    \\
  Avg. time & 79.92  & 243.42 &   584.24  & 1986.45 & 422.32    \\
  \bottomrule
  \end{tabular}
  }
\end{table}

\begin{table}
  \centering
  \caption{The performance comparison between the proposed method \proposed with three different initial remaining gene pool settings for \textit{iMM904}.
  Each gene deletion strategy was deemed successful if it achieved a minimum GR and PR of 0.001 or higher during GR maximization. The unit of elapsed time is second (s).}
  \label{table: e2_gene}
  \resizebox{0.99\linewidth}{!}{%
  \begin{tabular}{lcccc} 
  \toprule
  \begin{tabular}[c]{@{}l@{}}Initial remaining \\gene pool\end{tabular} & \begin{tabular}[c]{@{}c@{}}Predicted-$G_{\text{remain}}$ \\(Proposed)\end{tabular} & GE gene & $G_{\text{remain}}$   & RC gene \\ 
  \midrule
  Success rate     & 81/700    & 120/700           & 126/700        & 65/700 \\ 
  \midrule
  Time.Success = 1         & 22.89        & 189.90        & 47.82       & 31.93  \\
  Time.Success = 0         & 87.38           & 459.31      & 224.19       & 157.42\\
  Time.All cases       & 79.92           & 413.13       &  192.44       & 145.77\\
  \bottomrule
  \end{tabular}
  }
\end{table}

\textbf{For \textit{iML1515}:}
Table \ref{table: e3_compare} summarizes the performance comparison between the proposed method \proposed (using Predicted-$G_{\text{remain}}$ genes as the initial remaining gene pool) with GDLS, optGene, \gDel, and gMCSE in the context of targeting 991 non-extracellular space metabolites of \textit{iML1515}.
Regarding the success rate, \proposed achieved success in 507 out of 991 target metabolites, meeting the minimum GR and PR threshold of 0.001 or higher during GR maximization. 
GDLS, gMCSE, and optGene failed in all 991 cases, and \gDel succeeded in 508 out of 991 cases.
\proposed demonstrated outstanding performance with an average elapsed time of 431.75 seconds. 
GDLS had an average time of 244.58 seconds, optGene required significantly more extended time with an average of 2193.45 seconds, \gDel had an average time of 2541.23 seconds, and gMCSE had an average time of 1892.39 seconds.

Table \ref{table: e3_gene} summarizes the performance comparison between the proposed method \proposed with three different initial remaining gene pool settings. 
When utilizing Predicted-$G_{\text{remain}}$ genes as the initial remaining gene pool, the success rate in achieving the minimum GR and PR threshold of 0.001 or higher during GR maximization was 507 out of 991 instances. 
When using the GE gene as the initial remaining gene pool, the success rate was 508 out of 991 instances. 
When employing $G_{\text{remain}}$ genes as the initial remaining gene pool, the success rate remained at 508 out of 991 instances.
Additionally, when employing RC genes as the initial remaining gene pool, the success rate decreased to 178 out of 991 instances.
Examining the computational efficiency, the proposed method demonstrated varied performance regarding elapsed time for successful cases. 
For instances where success was achieved, the average elapsed time was 92.46 seconds with Predicted-$G_{\text{remain}}$ genes, 481.66 seconds with GE genes, 90.30 seconds with $G_{\text{remain}}$ genes, and 80.26 seconds with RC genes.
In contrast, for cases where success was not achieved, the average elapsed time was 787.17 seconds with Predicted-$G_{\text{remain}}$ genes, 4561.56 seconds with GE genes, 858.01 seconds with $G_{\text{remain}}$ genes, and 852.43 seconds with RC genes. 
Considering all cases, regardless of success or failure, the proposed method showed outstanding average elapsed times when using Predicted-$G_{\text{remain}}$ or $G_{\text{remain}}$ genes, with values of 431.75 and 463.70 seconds.

\begin{table}
  \centering
  \caption{The performance comparison between the proposed method \proposed (using Predicted-$G_{\text{remain}}$ genes as the initial remaining gene pool) with GDLS, optGene, \gDel, and gMCSE in the situation of targeting non-extracellular space metabolites of \textit{iML1515}. 
  Each gene deletion strategy was deemed successful if it achieved a minimum GR and PR of 0.001 or higher during GR maximization. The unit of elapsed time is second (s).}
  \label{table: e3_compare}
  \resizebox{0.99\linewidth}{!}{%
  \begin{tabular}{lccccc} 
  \toprule
  Method           & \begin{tabular}[c]{@{}c@{}}Proposed\\framework\end{tabular} & GDLS  & gMCSE  & optGene & gDel\_minRN   \\ 
  \midrule
  Success rate     & 507/991    & 0/991  &  0/991  & 0/991   & 508/991     \\
  Avg. time & 431.75  & 244.58 &  1892.39 & 2193.45 & 2541.23      \\
  \bottomrule
  \end{tabular}
  }
\end{table}

\begin{table}
  \centering
  \caption{The performance comparison between the proposed method \proposed with three different initial remaining gene pool settings for \textit{iML1515}.
  Each gene deletion strategy was deemed successful if it achieved a minimum GR and PR of 0.001 or higher during GR maximization. The unit of elapsed time is second (s).}
  \label{table: e3_gene}
  \resizebox{0.99\linewidth}{!}{%
  \begin{tabular}{lcccc} 
  \toprule
  \begin{tabular}[c]{@{}l@{}}Initial remaining \\gene pool\end{tabular} & \begin{tabular}[c]{@{}c@{}}Predicted-$G_{\text{remain}}$ \\(Proposed)\end{tabular} & GE gene & $G_{\text{remain}}$   & RC gene \\ 
  \midrule
  Success rate     & 507/991    & 508/991            & 508/991      &   178/991\\ 
  \midrule
  Time.Success = 1         & 92.46        & 481.66        & 90.30    &   80.26 \\
  Time.Success = 0         & 787.17           & 4561.56      & 858.01    &   852.43 \\
  Time.All cases       & 431.75           & 2470.15       & 463.70     &  713.74 \\
  \bottomrule
  \end{tabular}
  }
\end{table}

\begin{table}
  \centering
  \caption{Distribution of percentile ranks for target metabolites PRs across three metabolic models using DBgDel, with MetNetComp PRs as baselines.}
  \label{table: percentile}
  \resizebox{0.99\linewidth}{!}{%
  \begin{tabular}{lccc} 
  \toprule
  Percentile range & e\_coli\_core & iMM904  & iML1515  \\ 
  \midrule
  0\%                & 8.82\%        & 25.56\% & 0.91\%   \\
  0-30\%             & 38.23\%       & 6.67\%  & 1.21\%   \\
  30-60\%            & 0.00\%        & 4.44\%  & 31.72\%  \\
  60-90\%            & 0.00\%        & 1.11\%  & 0.91\%   \\
  \texttt{>}90\%               & 52.94\%       & 62.22\% & 65.26\%  \\
  \bottomrule
  \end{tabular}
  }
  \end{table}

\textbf{Comparison of PR Percentile Ranks:}
To further analyze the trade-offs in DBgDel and discuss the limitations of its resulting gene deletion solutions, we compare the PRs obtained by DBgDel with the PRs listed in MetNetComp for target metabolites in three models under study. 
The DBgDel PRs were derived from computational simulations of metabolic models using DBgDel gene deletion strategies.
The MetNetComp PRs were obtained from the MetNetComp database as a baseline for comparison.

Table \ref{table: percentile} summarizes the resulting distribution of percentile ranks for target metabolites PRs across three metabolic models using DBgDel.
Percentile ranks were calculated using MetNetComp PRs as baselines.
A percentile of 0\% indicates a zero production rate and, consequently, a failure to achieve growth-coupled production.
In the e\_coli\_core model, the largest proportion of target metabolites (52.94\%) fell within the \texttt{>}90\% percentile range, followed by 38.23\% in the 0-30\% range. 
Smaller proportions were observed in the 0\% (8.82\%) and the 30-60\% and 60-90\% ranges (both 0\%).
For the iMM904 model, the largest proportion of target metabolites (62.22\%) were in the \texttt{>}90\% range, followed by 25.56\% in the 0\% range. 
Smaller proportions were observed in the 0-30\% (6.67\%), 30-60\% (4.44\%), and 60-90\% (1.11\%) ranges.
In the iML1515 model, the \texttt{>}90\% range had the highest proportion of metabolites (65.26\%), followed by 31.72\% in the 30-60\% range. 
The 0\% range accounted for 0.91\%, and the 0-30\% and 60-90\% ranges accounted for 1.21\% and 0.91\%, respectively.

\section{Discussion and Conclusion}
\label{sec:discussion}
In this study, we developed a novel framework \proposed to calculate gene deletion strategies to achieve growth-coupled production in genome-scale metabolic models. 
\proposed integrates the gene deletion strategy database information to accelerate the algorithms' computational efficiency.
The experimental results above demonstrate the efficiency of the proposed framework, \proposed, in calculating gene deletion strategies for models of varying scales. 
Particularly, as the scale of the model increases, its computational efficiency advantages become more pronounced compared to other methods, with an average 6.1-fold acceleration. 
Notably, even when compared to the previously best-performing method, i.e., \gDel, \proposed achieves a 1.45/2.4/5.89 fold improvement in computational efficiency across three metabolic models, respectively.

Of greater significance, this enhancement in computational efficiency is accompanied by a respectable overall success rate in calculating gene deletion strategies for new target metabolites. 
Moreover, as the scale of the models increases, the discrepancy in success rates tends to diminish. 
Specifically, compared to the original \gDel, the gap in success rates for \proposed across three different models is only 10\%/5\%/0.1\%.
The results of the computational experiments affirm that the proposed framework, \proposed, effectively strikes a balance between success rate and computational efficiency.

We conducted a small-scale experiment using the e\_coli\_core model to evaluate the integration of a baseline method, gMCSE, into the proposed framework. The results showed a 32\% improvement in computational speed while the success rate decreased from 17.5\% to 12.5\%.
The robust trade-off between success rate and computation time was observed.
In this study, we configured the default settings of the gMCSE algorithm to ensure a logical setup, requiring that the designed strains produce metabolites during growth and meet the conditions for directional growth coupling. 
Additionally, we adhered to the framework’s default initial gene pool, the Predicted-$G_{remain}$ gene set. While the application of other algorithms and configurations to large-scale metabolic models remains unexplored, this example demonstrates that the proposed framework, with its initial gene pool, can be adapted to other gene deletion algorithms that properly account for GPR rules.

It is important to note that the implementation of GPR rules in this study differs from that in the gMCSE study described in \mbox{\cite{GPR}} and \mbox{\cite{schneider2020extended}}.
Our method directly implements Boolean functions, whereas the gMCSE method transforms GPR rules into a stoichiometric representation, resulting in some approximation.
Specifically, in the gMCSE study, when a reaction is not repressed by genes, its corresponding reaction rate cannot be 0. 
In contrast, in our study, even if a reaction is not repressed by genes, its reaction rate can still be 0.
As a result, gene deletions identified by gMCSE may not always be regarded as successful in our study particularly when a gene is associated with multiple reactions.
Additionally, cases where the algorithms reach the computational time limit without producing gene deletion results are also treated as unsuccessful.

Algorithms capable of calculating gene deletion strategies with a maximum number of gene knockouts are desirable. 
Such algorithms provide gene deletion strategies that reveal the core flow in the metabolic networks.
In other words, these algorithms help biological understanding of what is necessary for growth-coupled production, holding significant practical relevance in metabolic engineering.
Moreover, based on the resulting maximal gene deletion strategies, it is also possible to derive many other gene deletion strategies with fewer gene deletions \mbox{\cite{tamura2022trimming}}.
Nowadays, researchers have successfully utilized gene deletion algorithms to address real-world challenges in constructing mutant strains for achieving growth-coupled production of various target metabolites, with applications in biofuel production and industrial production of platform chemicals \mbox{\cite{trinh2008minimal, schneider2020extended,banerjee2020genome}}.

The experimental results using the RC genes indicate that STEP 1 of DBgDel effectively narrows the search space. 
The differences in success rates are 7\%, 9\%, and 33\% for \textit{e\_coli\_core}, \textit{iMM904}, and \textit{iML1515}, respectively, demonstrating that the effect becomes more pronounced as the model size increases.

The results of the comparison of PR percentile ranks demonstrate that, although DBgDel makes trade-offs between success rate and computational efficiency, its resulting gene deletion strategies still achieve high PRs (particularly in the \texttt{>}90\% percentile range) for a significant portion of target metabolites across all three metabolic models when compared to MetNetComp baselines.
This also implies that even compared to global optimization methods like RobustKnock \cite{tepper2010predicting}, —which cannot effectively compute large-scale gene deletions for genome-scale metabolic models—the iterative approach of DBgDel still delivers competitive local solutions in terms of resulting PRs.

The gene deletion strategies of MetNetComp achieve a local maximum in terms of the number of gene deletions. 
In contrast, the strategies generated by DBgDel are ensured to be neither local nor global maxima, as valid solutions may be obtained by adding or removing genes. 
Rather than optimizing the number of gene deletions or PR, DBgDel aims to identify acceptable gene deletion strategies efficiently within a short time.

In future work, we will explore reducing the search space for other gene deletion algorithms, demonstrating the broader applicability and potential performance gains of the DBgDel framework.
Additionally, we will study how to better extract and utilize information from the gene deletion database, such as achieving more flexible information extraction without relying on unified, handcrafted computing rules.

\section{ACKNOWLEDGMENTS}

This work was supported in part by JSPS, KAKENHI under Grant \#20H04242, and JST SPRING, Grant Number JPMJSP2110.

\bibliographystyle{naturemag}
\bibliography{Bibliography}
\end{document}